

\input harvmac
\def\eg{{\it e.g.\/}}
\def\t{\theta}

\def\np#1#2#3{Nucl. Phys. {\bf #1} (#2) #3}
\def\pl#1#2#3{Phys. Lett. {\bf #1} (#2) #3}
\def\prl#1#2#3{Phys. Rev. Lett. {\bf #1} (#2) #3}
\def\pr#1#2#3{Phys. Rev. {\bf #1} (#2) #3}

\def\blankref#1#2#3{   {\bf #1} (#2) #3}

\Title{\vbox{\hbox{SSCL--PP--167}\hbox{BROWN HET--882}\hbox{hep-ph/9211216}}}
{On Constraints for Heavy-Meson Form Factors}

\centerline{Benjam\'\i n Grinstein\footnote{$^{\star}$}
{\tt grinstein@sscvx1.ssc.gov\ or @sscvx1.bitnet}
and Paul F. Mende\footnote{$^{\star\star}$}
{\tt mende@het.brown.edu}}
\bigskip\centerline{\it $^\star$Superconducting Super Collider Laboratory,
2550 Beckleymeade Ave, Dallas, Texas 75237}
\centerline{\it $^{\star\star}$Department of Physics, Brown University,
Providence, Rhode Island 02912}

\bigskip
\bigskip

We examine the recent work of de~Rafael and Taron where model-independent
bounds on the Isgur-Wise function are presented.
We first argue that the bounds cannot
hold in as much generality as implied.
We show that the effects of resonances
omitted in their discussion
(such as heavy-heavy ``onium'' states below threshold)
modify the bound.
The resulting bound is much weaker but may be useful where
the size of the additional contribution may be computed or estimated.

\Date{November 1992}
\newsec{ Introduction}
In the theoretical limit of very large charm and bottom masses,
$\Lambda_{\rm QCD}\ll m_c\ll m_b$, approximate
symmetries\ref\isgurwise{N. Isgur and M.B. Wise, \pl{B232}{1989}{113};
\pl{B237}{1990}{527}}\  of QCD allow
one to express all six form factors relevant to semileptonic decays of
$B$ mesons into $D$ and $D^*$ mesons in terms of the $B$-number
form
factor of the $B$ meson, $F(q^2)$:
\eqn\bff{
\vev{\bar B(p') | \bar b\gamma_\mu b | B(p)} = F(q^2) (p + p')_\mu
}
where $q=p-p'$. $B$-number conservation implies $F(0)=1$. Thus the
six
form factors for $\bar B\to D^{(*)}\ell\nu$ are known, in this limit, at
one kinematic point. This is remarkable because it allows in principle
for the extraction of the mixing angle $V_{cb}$ from experiment
without
the need for a phenomenological model of the hadrons involved. In
practice, however, the differential decay rate vanishes at the
kinematic point where the form factors are predicted,
so an extrapolation is necessary.
The extracted value of the mixing angle depends on the nature
of the extrapolation. Different models suggest different functional
dependence of $F(q^2)$, and we are stuck again with a model
dependent extraction of the mixing angle.

De~Rafael and Taron\ref\DT{E. de~Rafael and J.~Taron,
\pl{B282}{1992}{215}}\ have recently
argued that the form factor $F(q^2)$ must
satisfy some rather restrictive model-independent conditions in the
form of bounds:
\eqn\boundone{
F_-(q^2) < F(q^2) < F_+(q^2)
}
where $F_\pm(q^2)$ are known functions which depend on $F(0)$.
If these bounds were correct, they would rule out
some models which have been used to
extract $V_{c b}$ from experiment, making the determination of this
mixing angle far more precise.

Our discussion is organized as follows.
In section~2 we review the
derivation of Ref.~\DT.
In section~3 we note that their conclusion conflicts
with the known physical behavior when applied to some other systems.
Section~4 identifies the omission of Ref.~\DT:
part of the states contributing to the analytic structure of $F(q^2)$
were neglected.
We note why this omission is more severe for heavy-heavy quark states.
We show how the bound may be modified to include additional
contributions.
The resulting bound is useful, however, only if one can estimate
or compute the couplings of additional states.
We explain how this works in models, and consider in
particular the case of the 't~Hooft model.
In the last section we suggest some avenues of research that may
yield somewhat model dependent but useful bounds.

\newsec{ Derivation of the Bound}
Our derivation of the bounds follows closely that of de~Rafael and
Taron,
except that we avoid strictly
taking the limit $m_b\to\infty$ and we work with
arbitrary dimensions of spacetime, $D$.
Consider the two point function of the heavy current $J_\mu=\bar
b\gamma_\mu b$
\eqn\tpf
{
(q_\mu q_\nu-q^2g_{\mu\nu})\Pi(q^2)=
i \int d^D\!x e^{i q x}\vev{{\rm T} J_\mu(x) J_\nu(0)}.
}
In QCD, for $D<6$ it satisfies a once subtracted dispersion relation:
\eqn\osdr{
\chi(Q^2)=-\left.{{\partial\Pi}\over{\partial q^2}}\right|_{q^2=-Q^2}=
{1\over\pi}\int_0^\infty dt \, {{{\rm Im}\,\Pi(t)}\over{(t+Q^2)^2}}.
}
The absorptive part, ${\rm Im}\,\Pi(q^2)$, corresponds to all the
contributions to the right-hand side of Eq.~\tpf\ with a real (on shell)
state
between the two current insertions. This is a sum of positive definite
terms, so one can obtain strict inequalities by neglecting some
contributions to the sum. Concentrating on the term with
intermediate
states of $B\bar B$ pairs, one finds
\eqn\pigf{
{1\over\pi}{\rm Im}\,\Pi(q^2)
\ge {1\over{(4\pi)^{(D-1)/2}\Gamma({{D+1}\over2})}}
{{(q^2/4-M_B^2)^{(D-1)/2}}\over{(q^2)^{3/2}}}  |F(q^2)|^2
\theta(q^2/4-M_B^2),
}
where the function $F$ is the analytic continuation of the form factor
in Eq.~\bff\ for spacelike momentum transfer. The dispersion
relation
can be used to extend this inequality into the timelike region:
\eqn\chigf{
\chi(Q^2) \ge
{1\over{(4\pi)^{(D-1)/2}\Gamma({{D+1}\over2})}}
\int_{4M_B^2}^\infty \!\!\!dt\,\,\,
{{(t/4-M_B^2)^{(D-1)/2}}\over{t^{3/2}(t+Q^2)^2}} \, |F(t)|^2.
}
For large $b$-quark mass the two point function can be reliably
computed
from perturbative QCD on the timelike region, and in particular at
$q^2=0$. With $N_c$ colors, it is easy to verify that, to leading order in
$\alpha_s$,
\eqn\chiqcd{
\chi(Q^2) = {{2N_c}\over{(2\pi)^{D/2}}}\Gamma(3-D/2)
\int_0^1 dx\, x^2(1-x)^2 (m_b^2+x(1-x)Q^2)^{D/2-3}.
}
Setting $Q=0$, changing variables to $y=t/4M_B^2$
and recalling that $M_B/m_b=1 + {\cal O}(\Lambda_{\rm QCD}/m_b)$,
one
has, to leading order in $\Lambda_{\rm QCD}/m_b$,
\eqn\onegf{
\int_1^\infty dy\, k(y) |F|^2 \le 1,
}
where we have introduced
\eqn\kdefd{
k(y)= C y^{-7/2} (y-1)^{(D-1)/2}
}
and
\eqn\cdefd{
C= { {15\sqrt{\pi}} \over {2^{D/2+4} N_c \Gamma({{D+1}\over2})
\Gamma({{6-D}\over2})}}.
}

\nref\meiman{
	N. N. Meiman, Sov. Phys. JETP \blankref{17}{1963}{830}\semi
	S. Okubo and I. Fushih, \pr{D4}{1971}{2020}\semi
	V. Singh and A.K. Raina, Fortschritte der Physik
	\blankref{27}{1979}{561}}%
\nref\bourrely{
	C. Bourrely, B. Machet and E. de Rafael, \np{B189}{1981}{157}}

Next we argue\refs{\meiman,\bourrely}\ that any function $F$ satisfying
\onegf\ and analytic except
for a cut on the real axis for $y\ge1$ is bounded as in \boundone, and we will
explicitly construct the functions $F_\pm$ for any $D<6$. To this end we map
the complex $y$-plane onto the unit disk $|z|\le1$ by the transformation
\eqn\ytoz{
\sqrt{y-1}=i{{1+z}\over{1-z}}
}
The cut $y\ge1$ is mapped   into the unit circle
$z=e^{i\theta}$. (The two edges of the cut are mapped into the upper and
lower semicircles) The timelike region $y\le0$ is mapped into the
segment of the real axis $0\le z<1$. In terms of this new variable
the
inequality \onegf\ is
\eqn\onegfb{
{1\over2}\int_0^{2\pi} d\t\, w(\t) |F|^2 <1,
}
where
\eqn\ktow{
\eqalign{
w(\t) &= k(y(\t)) {{dy}\over{d\t}}\cr
      &=C \cos^D\theta/2\,\sin^{5-D}\theta/2.\cr}
      }
As we will see shortly, the derivation of the inequality uses a
function
$\phi(z)$
analytic in $|z|<1$ such that $|\phi(e^{i\theta})|^2=w(\theta)$. The
construction of such a function is not unique ---one may always
multiply
one such function by a power of $z$. We make the choice that
reproduces
the result of de~Rafael and Taron:
\eqn\phidefd{
\phi(z)=C^{1/2} 2^{-5/2} (1+z)^{D/2} (1-z)^{(5-D)/2}.
}

With ref.~\bourrely, let us define an inner product on the space of complex
functions of a real variable $\theta$, with $0\le\theta<2\pi$, by
\eqn\ipdefd{
(f,g)\equiv{1\over2\pi}\int_0^{2\pi}d\theta\, f^*(\theta)g(\theta).
}
Next let
\eqn\fis{
\eqalign{
f_1(\t) &= \phi(e^{i\t}) F(e^{i\t})\cr
f_2(\t) &= {1\over{1-\bar z_0 e^{i\t}}}\cr
f_3(\t) &=1\cr}
}
With this, we have
\eqn\Idefd{
I\equiv (f_1,f_1)={1\over2\pi}\int_0^{2\pi} d\t\, w(\t)
|F|^2\le{1\over\pi}
}
Using Cauchy's theorem we can evaluate the other inner products.
For
example, with a contour C given by the unit circle $|z|=1$,
\eqn\foneftwo{
g(z_0)  ={1\over2\pi i}\int_C dz\, {g(z)\over z-z_0}
        = {1\over2\pi} \int_0^{2\pi} d\t\, {g(e^{i\t})\over 1-z_0
       e^{-i\t}}=(f_2,f_1)
       }
 From the positivity of the inner product we have that the $3\times 3$
matrix $(f_i,f_j)$ has positive determinant. This gives
\eqn\posdet{
{\rm det}\pmatrix{I&g(z_0)&g(0)\cr
                  g^*(z_0)&{1\over 1-|z_0|^2}&1\cr
                  g^*(0) & 1 &1\cr}
          \ge0.
          }
The inequality \boundone\
$$F_-(q^2) < F(q^2) < F_+(q^2)
$$
 follows, with
\eqn\fpmdefd{
F_\pm(z)= {g(0)\over\phi(z)}\left[1\pm{|z|\over\sqrt{1-
|z|^2}}\sqrt{{1\over\pi|g(0)|^2}-1}\,\right]
}
The result can be used to obtain a bound on the charge radius,
\eqn\chrad{
\langle r^2 \rangle =
2(D-1){dF\over dq^2}(0)\le{D-1\over8M_B^2}\left[D-{5\over2}+\sqrt{{2^5\over
\pi C}-
1}\right]
}

De~Rafael and Taron phrase their result as a bound on the Isgur-Wise
function, by taking the infinite mass limit literally. We refrain from
taking the limit $m_b=\infty$ since for $vv' >1$ this sends the form factor to
zero. The Isgur-Wise function is not a physical
observable,
\eg, it depends on the renormalization point $\mu$. In fact, the form
factor $F(q^2)$ is related to the Isgur-Wise function $\xi(vv',\mu)$
through
\eqn\fiw{
F(q^2=2M_B^2(1-vv')) =
\left({\alpha_s(m_b)\over\alpha_s(\mu)}\right)^{a_L(vv')}
\xi(vv',\mu)
}
in the limit $m_b\to\infty$. In this limit only the leading log term
needs be retained. In Eq.~\fiw\ the anomalous dimension is
\eqn\asubl{
a_L(vv')={8\over33-2n_f}[vv' r(vv') -1]
}
where $n_f=4$ is the number of active flavors and
\eqn\rdefd{
r(x) \equiv { 1 \over \sqrt{x^2 -1}} \ln \left(x + \sqrt{x^2-1}\right).
}
Notice that, for $vv'>1$, the factor
$({\alpha_s(m_b)/\alpha_s(\mu)})^{a_L(vv')}$ vanishes as $m_b\to\infty$.

\newsec{ Trouble on the horizon}

De~Rafael and Taron point out that some models of the form factor
violate the bound \boundone; they view this as a shortcoming of
such models.

One can turn this sort of observation around to cast
doubt on the validity of the bound by trying to apply it first to
simpler physical cases than the intricate
heavy-light systems which are of primary interest.
In particular, one may argue that the bound Eq.~\chrad\ cannot
be correct because
the charge radius of hydrogen atom-like wave-functions for the $B$
meson
must become arbitrarily large in the weakly bound case of the
small coupling limit.

Isgur\ref\private{Private communication}\ has pointed out how to refine
this objection further. The derivation
above holds not just for the form factor of the $B$ meson, but also
for that of any particle that can be pair produced from the vacuum by
the current $J_\mu$. Thus, the bound must hold for the form factor of
the $B_c$ meson, and even for that of the $\eta_b$. Now, these
are well known to be well described by  weakly bound
non-relativistic wave-functions. Moreover, the charge radius is
enormous
compared to the Compton wavelength of the $b$-quark, so that the
violation of the bound is rather dramatic (as opposed to the rather
marginal violation in the case of heavy-light systems).

Clearly there is trouble, and
a satisfactory explanation of what is wrong with the bound
should account for the different behavior of form factors of
heavy-light and heavy-heavy systems.

\newsec{ The demise of the bound.}

Let us analyze in detail the assumptions and manipulations that lead to the
form factor bound Eq.~\boundone.
A number of physical considerations lead to the
intermediate inequality~\onegf, which plays a crucial role.
Is this inequality valid?
As far as we can tell there is nothing wrong with the
derivation that
leads to it. In fact, the inequality is rather weak because one expects
the
absorptive part of the two point function, ${\rm Im} \Pi$, to be
dominated by
resonant intermediate states ($\Upsilon$, $\Upsilon'$, etc).
Is the perturbative QCD calculation of $\chi$ reliable at $Q^2=0$ rather than
in the more familiar limit $Q\to\infty$?
We believe it is. The onset of the physical
cut is at spacelike momentum and scales with the square of the
large mass $m_b$.
Since this is much larger than the hadronic scale one expects QCD
to be a very good
approximation for timelike (and lightlike) momentum.

Consider therefore the second half of the derivation
(following Eq.~\onegfb) in which a positivity condition was derived
for the function~$F(y)$.
One of the key assumptions about the analytic structure
of~$F$ --- that $F(q^2)$ have no singularities below the $B$-threshold
cut at $q^2 \ge 4M_B^2$ --- is in fact violated for good physical reason.
This assumption is used to deduce
$(f_2,f_1)=g(z_0)$ in Eq.~\foneftwo,
and in computing $(f_3,f_1)=g(0)$.
Without this condition, the contours cannot be suitably deformed.
We will argue that the form factor~$F(q^2)$
has singularities for $q^2\le4M_B^2$, and that these modify the
bounds in~\boundone\ even in the large $m_b$ limit.

The singularities we are referring to correspond to processes in which an
`onium' state $\Upsilon_n$ is created by the current out of the vacuum with
amplitude $f_n$, and then
couples to the $B\bar B$ pair with amplitude $g_{nB\bar B}$.
For simplicity, let us take these states to be stable so they contribute
poles rather than cuts to the form factor.
(The modification replacing poles by cuts and residues by integrals over
discontinuities is straightforward and described below.)
Since several of these states may lie
below threshold, $F(q^2)$ has a term of the form
\eqn\polesone{
\sum_n { f_n g_{nB\bar B}\over q^2-M_n^2}.
}
When the $q^2$ plane is mapped into the circle $|y|\le 1$, these
poles land in the interior of the circle and hence in the interior
of the contour of Eq.~\foneftwo.
Now the naive scaling of these quantities for large $m_b$ is a factor
of
$\sqrt{m_b}$ for each heavy meson in an amplitude, so
\eqn\fgscaling{
f_n\sim m_b^{1/2} \qquad  {\rm and} \qquad g_{nB\bar B}\sim
m_b^{3/2},
}
and of course $M_n\sim m_b$. Therefore the
contribution~\polesone\ can
be rewritten as
\eqn\polestwo{
\sum_n {\tilde f_n \tilde g_{nB\bar B}\over y-\tilde M^2_n},
}
where $f_n= (2M_B)^{1/2}\tilde f_n$, $g_{nB\bar B}=(2M_B)^{3/2}
\tilde g_{nB\bar B}$, $M_n^2=4M_B^2 \tilde M_n^2$ and, as above,
$y=q^2/4M_B^2$. Clearly this contribution cannot be ignored even in
the large mass limit.

When these poles are taken into account Eq.~\foneftwo\ is no longer
valid.
Instead they generate an additional contribution:
\eqn\foneftwob{
(f_2,f_1)=g(z_0)+\tilde g(z_0),
}
The explicit form of $\tilde g$ for the case of pole singularities is
\eqn\tildg{
\tilde g(z)= \sum_n\left(
{\tilde f_n\tilde g_{nB\bar B}(1-z_{n-})^2 \phi(z_{n-})
\over\tilde M_n^2(z_{n-}-z_{n+})}\right)
{1\over z-z_{n-}},
}
where $z_{n\pm}$ are the roots of $z^2\tilde M_n^2-2z\tilde M_n^2+4z+\tilde
M_n^2=0$, with $|z_{n-}|<1$.

A new bound is still obtained in this fashion, but without {\it a priori\/}
knowledge of the quantities $f_n$, $g_{nB\bar B}$ and $M_n$ it is
hardly very useful.
To obtain interesting constraints, one needs to first estimate
the size of these contributions and second, to work with a system
in which they can hopefully be shown to be small.

Figure~1\nfig\onlyone{$z$-plane with the
disc $|z|\le1$ corresponding to the complex $q^2=4M_B^2y$ plane as per
Eq.~\ytoz. The heavy line depicts the cut from some branch point below
threshold
out to infinity. } shows the $z$-plane with the
disc $|z|\le1$ corresponding to the complex $q^2=4M_B^2y$ plane as per
Eq.~\ytoz. The heavy line depicts the cut from some branch point below
threshold
out to infinity.  It is clear from the figure that the angular integral
in Eq.~\foneftwo\ and the contour integral cannot be made to agree, since the
contour cannot be taken to cross the cut. An appropriate
choice of contour differs from the angular integral by an integral over
the discontinuity across the cut for $|z|\le1$. Hence, again, Eq.~\foneftwo\
must be replaced as in Eq.~\foneftwob, $(f_2,f_1)=g(z_0)+\tilde g(z_0)$, where
now, in more generality,
\eqn\foneftwob{
\tilde g(z_0)=-\int_{-1}^{z_{\Upsilon}} dz \quad{\rm Im}\, {g(z)\over z-z_0}~.
}
Here $z_{\Upsilon}$ stands for the image under~\ytoz\ of the location of the
first branch point, $y_\Upsilon=M_\Upsilon^2/4M_B^2$ (discontinuities
associated with light particles, \eg, $\pi\pi$, are present but decouple in the
large $m_b$ limit).

We can now understand why the putative bound Eq.~\chrad\ was more
severely violated for
heavy-heavy mesons than for heavy-light mesons.
As above we can phrase
the question in terms of how large the slope of the form factor at
$q^2=0$, the charge radius, may be.
In ref.~\ref\bobj{
	R.L. Jaffe,  \pl{245B}{1990}{221}}\
and ref.~\ref\jm{
	R.L.~Jaffe and P.F.~Mende,
	\np{B369}{1992}{189}}
it was shown that the anomalously large radius of a heavy-heavy
QCD bound state
arises from the presence of many closely spaced singularities with
significant residues.
It is because the effect of such states was missed
that de~Rafael and Taron obtained an overly stringent bound
on the charge radius of the heavy-heavy states.

Furthermore, if one of these quarks is taken to be increasingly light,
some of these singularities migrate above
threshold until, for a heavy-light system, one is left with
only a few resonances
below threshold (recall that the $\Upsilon^{(\prime\prime\prime)}$ is
already above threshold). Therefore, the violation of the naive
bound Eq.~\boundone\
is less severe in these cases.
Finally, we note that
resonances continue to migrate above threshold for light-light systems,
so the objections do not apply to the arguments of ref.~\bourrely\ (although
the
applicability of perturbative QCD in that case may cast a doubt on
those
results).

Finally, we have examined all of these considerations in a relevant
solvable toy model.
We investigated the validity of these bounds in $1+1$ dimensions
in the large-$N_c$ limit.
This is an ideal testing ground for these issues since, on the one hand
the quantities $g_{nB\bar B}$ and indeed the full singularity
structure of the form factors can be computed exactly; and on the other
hand much of the dynamics does in fact parallel 3+1 dimensional
heavy quark physics in regimes where they can be compared.
It is well known that this is a good approximation for
$N_c\ge4$. The bound in eq.~\boundone, now with $D=2$, holds for
heavy-light systems provided $N_c\ge4$, and is violated for
$N_c\le3$.
(This can be seen in Fig.~2 of ref.~\ref\gm{B.~Grinstein
and P.F.~Mende, \prl {\bf 69} {1992} 1018}.)
The bound is violated for heavy-heavy systems, even for rather
large
values of $N_c$, recovering the observation above (but now without
recourse to non-relativistic hadronic models)\jm.

In this model\ref\thooft{G.~'t~Hooft,
        \np{\bf B75}{1974}{461}\semi
C.~G.~Callan, N.~Coote, and D.~J.~Gross,
        \prl {\bf D13}{1976}{1649}\semi
M.~B.~Einhorn,
        \prl {\bf D14}{1976}{3451}}\
 the
absorptive part is saturated by resonances, so it has the form of an
infinite
sum over single poles on the positive $q^2$ axis, so that
 \eqn\chilargen{
\chi(Q^2)= \sum_n {f_n^2\over(Q^2+m_n^2)^2}.
}
The coefficients $f_n$ and the masses $m_n$ can be calculated
numerically. We
have checked, numerically, that for $Q^2\ge0$ the perturbative
expression,
Eq.~\chiqcd\ with $D=2$, is in excellent agreement with the sum in
Eq.~\chilargen.

\newsec{ Conclusions}

We have shown the form factor bounds on $F(q^2)$ for
heavy-light mesons claimed by de~Rafael and Taron
do not hold because their derivation neglects
contributions to the form factor below the $B$-threshold, for
$q^2<4M_B^2$.

While the model-independent bound on form factors is lost, we can still derive
a bound on form factors if we are willing to use a model to estimate the
discontinuity across the cut in $F(q^2)$ for $q^2<4M_B^2$.
For example, a good
approximation may be a sum of poles corresponding to the upsilon resonances
below threshold, as in Eq.~\polestwo. If one can either measure or estimate
the constants $f_n$ and $g_{nB\bar B}$ that determine the residues, then a
bound on the form factor is obtained.  By including the effect of these very
resonances on the right hand side of \pigf\ an interestingly tight bound may
follow.  This new `model' of the form factor seems worth pursuing.

It is also worth pointing out that nontrivial information about the form
factors for the decay of heavy-light mesons into light-light mesons may
possibly be
obtained through similar methods. The argument is similar to that
presented above, but now one considers the two-point function of a
heavy-light current, $J_\mu=\bar q \gamma_\mu Q$, and its hermitian
conjugate.
The disadvantage in this case is that one no longer has knowledge, {\it a
priori}, of the value of the form factor at one value of $q^2$. On the other
hand, there is only one resonance below threshold, $M_{B^*}^2<q^2_{\rm
thresh}=(M_B+M_\pi)^2$. Thus, if one is willing to use, say, the constituent
quark model\foot{Alternatively, some or all of these quantities may
eventually be obtained from Monte Carlo simulations of lattice QCD, but a
computation of the form factor over this wide momentum range may prove
difficult on the lattice.}\ to calculate both $F$ at $q^2_{\rm max}=(M_B-
M_\pi)^2$, and the two constants $f_{B^*}$ and $g_{B^*B\pi}$, then one may
obtain a bound on the form factor for all momentum in the physical region for
the decay $B\to\pi e\nu$.  With any luck this bound will be tight enough that
the form factor is determined to good accuracy.

\bigskip

{\it Acknowledgments.\/} We are grateful to A. Falk, M. Luke and M.
Wise for communicating their results of a similar investigation prior
to publication\ref\flw{
A. Falk, M. Luke and M. Wise, SLAC-PUB-5956, UCSD/PTH 92-35,
CALT-68-1830, hep-ph@xxx/9211222}.
We would like to thank N.~Isgur for many
interesting discussions.
The research of B.G. is funded in part by
the
Alfred P. Sloan Foundation. This work is supported in part by the
Department of Energy under contracts DE--AC35--89ER40486
and DE-AC02-76-ER03130.

\listrefs

\listfigs

\bye